\documentclass[aps,twocolumn,superscriptaddress,hyperref]{revtex4}
\usepackage{graphicx,psfrag,amsmath,times,hhline}
\begin{document}

\title{Classification of Periodic, Chaotic and Random Sequences using NSRPS Complexity Measure }
\author{Karthi Balasubramanian, Gayathri R Prabhu, Lakshmipriya VK, Maneesha Krishnan, Praveena R and Nithin Nagaraj}
\affiliation{Department of Electronics and Communications, Amrita
School of Engineering, Amrita Vishwa Vidyapeetham, Amritapuri
Campus, Clappana P.O., Kollam, Kerala - 690 525.}

\begin{abstract}

Data compression algorithms are generally perceived as being of interest for data communication and storage purposes only. However, their use in the field of data classification and analysis is also of equal importance. Automatic data classification and analysis finds use in varied fields like bioinformatics, language and sequence recognition and authorship attribution. Different complexity measures proposed in literature like Shannon entropy, Relative entropy, Kolmogrov and Algorithmic complexity have drawbacks that make these methods ineffective in analyzing short sequences that are typical in population dynamics and other fields.

In this paper, we study Non-Sequential Recursive Pair Substitution (NSRPS), a lossless compression algorithm first proposed by Ebeling {\it et al.} [Math. Biosc. 52, 1980] and Jim\'{e}nez-Monta\~{n}o {\it et al.} [arXiv:cond-mat/0204134, 2002]). Using this algorithm, a new complexity measure was recently proposed  (Nagaraj {\it et al.} [arXiv:nlin.CD/1101.4341v1, 2011]). In this work, we use NSRPS complexity measure for analyzing and classifying symbolic sequences generated by 1D chaotic dynamical systems. Even with learning data-sets of length as small as 25 and test data-sets of length as small as 10, NSRPS measure is able to accurately classify the test sequence as periodic, chaotic or random. For such short data lengths, methods which use entropy measure and traditional lossless compression algorithm like LZ77 [A.Lempel and J.Ziv, IEEE Trans. Inform. Theory {\bf 22}, 75 (1976)] (used for instance by {\it Gzip}, {\it Winzip} etc.) fails.


\end{abstract}

\maketitle

\section{Introduction}
%
One of the major scientific advances made in the last century was the discovery that information content of a message can be objectively measured and quantified. This measure of information called the Shannon entropy ($H(X)$) plays a major role in both theoretical and practical aspects of information theory. Entropy estimation is very useful since it is a direct measure of the amount of compressibility that can be achieved for the given sequence. As per Shannon's Lossless Coding theorem~\cite{Shannon48}, entropy $H(X)$ gives the bound on the maximum possible lossless compression that can be achieved by a lossless compression algorithm. A truly random sequence would have maximum entropy and would thus be uncompressible. Entropy serves as an useful indicator of {\it complexity} of the data-set (higher the complexity, lesser would be the ability to compress).

Shannon's entropy measure \cite{Shannon48} is given by the following expression:
\begin{equation}
H(X) = -\sum_{i=1}^{M} p_i \log_2(p_i)~~~~\text{bits/symbol},
\end{equation}
where $X$ is the symbolic sequence with $M$ distinct symbols and $p_i$ is the
probability of the $i$-th symbol for a block-size of one. Block-size refers to the number of
input symbols taken together to compute the probability mass function.

Entropy may be a good measure but its estimation is not trivial. To calculate entropy of a time series, it needs to be converted to a symbolic sequence which requires the use of partition. The choice of a partition greatly affects the entropy value~\cite{Ebeling}. Also, noise is another factor that increases entropy. The length of the sequence also has a significant impact in entropy estimation. For most entropy estimation algorithms, accurate estimation is possible only for long sequences. For short sequences, the above equation is not a very good estimate of entropy and we resort to the use of lossless data compression algorithms instead. In this regard, Benedetto {\it et al.}~\cite{LangTree} refer to how compression algorithms like Lempel-Ziv (used by {\it Gzip, zip, WinZip}) may be used to estimate the complexity of a sequence. When LZ-77 compression algorithm encodes a sequence of length $L$ with entropy $H$,  into a zipped file of length $L_{z}$, then $\lim_{L\rightarrow\infty}(\frac{L_{z}}{L})\rightarrow H$. Using this, they define relative entropy and find similarities between sequences for automatic identification of unknown sequences. Using such a measure, Benedetto {\it et al.}~\cite{LangTree} have been quite successful in applications involving long sequences for purposes of language recognition, authorship attribution and language classification. While such a measure may work for long sequences, it may not be suitable for chaotic sequences which are short and noisy.

Due to these difficulties faced in entropy estimation, various other complexity measures such as Lyapunov exponent, Kolmogorov complexity, Algorithmic complexity, Grammar complexity and others have been proposed in literature~\cite{Measures}. One measure of complexity of interest to us is Grammar complexity, introduced by Ebeling {\it et al.} in~\cite{nsrps1}. The idea behind this measure is to compress a sequence and take the length of the compressed sequence as a measure of the complexity of the sequence.

The question to probe is, whether Shannon entropy and measures based on lossless compression algorithms are effective in all situations. Nagaraj {\it et al.}~\cite{NovelMeasure} have shown with an example that Shannon entropy might not be an effective complexity measure for certain sequences of smaller lengths and have proposed a new measure based on the NSRPS algorithm. In this paper, we investigate the use of this proposed NSRPS complexity measure for automatic classification of sequences generated by chaotic maps. 

This paper is organized as follows: Section II deals with the NSRPS algorithm and NSRPS measure as proposed in~\cite{NovelMeasure}. Section III deals with the problem of automatic classification and identification of periodic, chaotic and random sequences. Subsequently, in section IV, test results on various sequences are provided and the paper concludes in section V with indications of future research directions. 

\section{NSRPS Algorithm and NSRPS Measure}

Non-sequential Recursive Pair Substitution (NSRPS) was first proposed by Ebeling {\it et
al.}~\cite{nsrps1}. It was further improved by Jim\'{e}nez-Monta\~{n}o
{\it et al.}~\cite{nsrps2} and subsequently shown to be
optimal~\cite{nsrps3}. Grassberger employed NSPRS to estimate Entropy
of written English~\cite{Grassberger}. The algorithm (as described in ~\cite{NovelMeasure}) proceeds as follows. Let the
original sequence be called $X$. At the first iteration,
that pair of symbols which has maximum number of occurrences is replaced with a new symbol. For
example, the input sequence `$11010010$' is transformed into
`$12202$' since the pair `$10$' has maximum number of occurrences
compared to other pairs (`$00$', `$01$' and `$11$'). In the second
iteration, `$12202$' is transformed to `$3202$' since `$12$' has
maximum frequency (in fact all pairs are equally likely). The algorithm proceeds in this fashion until the
length of the string is 1 (at which stage there is no pair to
substitute and the algorithm halts). In this example, the algorithm transforms the input sequence `$11010010$' $\mapsto$ `$12202$' $\mapsto$ `$3202$' $\mapsto$ `$402$'  $\mapsto$ `$52$' $\mapsto$ `$6$'.

In ~\cite{NovelMeasure}, it is observed that the number of iterations needed to transform the input sequence to a constant sequence is a reasonable measure of {\it complexity} of the sequence.  For a constant sequence, the quantity `$entropy \times length$' is zero since entropy is zero (irrespective of length of the sequence).  We record the number of iterations as the complexity measure $N$. In the same paper, it was shown that this measure was successful in distinguishing sequences from chaotic dynamical systems such as the Logistic map for different values of the bifurcation parameter. It was also demonstrated that the measure has a high correlation with the Lyapunov exponent of the time series, even for very short data-lengths (as low as 50). Furthermore NSRPS measure is very easy to compute.

In the next section, NSRPS measure will be used to automatically classify a sequence of unknown complexity given several sequences of known complexities.
\section{Data classification and Identification}
The process of data classification and identification begins with first quantifying the information contained in a sequence. Having objectively quantified the information content, the concept of remoteness between pairs of sequences based on relative information content is defined and used as a distance metric between two sequences.
Our goal is to use this distance metric and device a method for automatically identifying unknown sequences of different complexities.
The problem at hand may be stated as follows: We are given known sequences (which we use as learning sequences) of four different complexities of length $L$$_{1}$ and are then provided with an unknown test-sequence $x$ of length $L$$_{2}$. We suspect that the test-sequence is one among the four known sequences and our aim is to determine the origin of the test-sequence.

A similar kind of problem is solved by Benedetto {\it et al.}~\cite{LangTree} in the field of automatic language recognition. Given two sequences $A$ and $B$, long as well as short sub-sequences are extracted from $A$ and $B$. An unknown short sequence $x$ which is to be classified is given. New sequences are formed by appending $x$ with long sequences of $A$ and $B$. Both the original and the new sequence are zipped using {\it Gzip} and the differences in their compressed lengths are noted. Let us denote the differences as $D_1$ and $D_2$ where $D_1$ =  $L_{A+x} - L_{A}$ and $D_2 = L_{B+x} - L_{B}$. The units of these quantities is bytes. $x$ is classified as belonging to that sequence which yields the minimum difference. They have been able to successfully recognize languages using an initial learning sequence of length 1-15 kilobytes and a test (unknown) sequence $x$ of length 20 characters.

There are a few drawbacks of the above approach, to name a few:
\begin{itemize}
\item Usage of entropy requires learning.  Hence the initial learning sequence has to be of very large length. This might not be practically possible for certain time series (for e.g. insect population counts).
\item Entropy measure is suitable only for certain data types. For data generated from chaotic dynamical systems, entropy might not be a good complexity measure.
\end{itemize}

We shall demonstrate that the above method of Benedetto is not suitable for short sequences obtained from chaotic dynamical systems. Instead, we find that NSRPS measure is more successful for classifying sequences from such systems.

\section{Results}

In this section, we present results pertaining to data from chaotic dynamical systems. We use the popular logistic map $y = ax(1-x)$ with different values of bifurcation parameter $a$ (3.83 and 4) to generate long time series of different complexities. Data produced with $a=3.83$ is periodic, while $a=4$ exhibits chaos as characterized by a positive lyapunov exponent. We use a partition with four bins of equal size to produce symbolic sequences from these time series. We also generate uniformly distributed random sequences so that we have three different learning sequences with different complexities. We also create three short sequences in a similar fashion. Thus we have a long and a short sequence for each of the three cases -- periodic ($a=3.83$), chaotic ($a=4.0$) and random (uniform). Our task is to automatically classify the short sequences given the long ones.

We shall use the following three measures of complexity and compare results to determine the best one:
\begin {itemize}
\item First order entropy measure $H(X)= -\sum_{i=1}^{n}{p(x_{i})\log_2(p(x_i))}$ bits/symbol~\footnote{For calculating first order entropy we take one symbol at a time (block-size is set to 1).}
\item {\it Gzip} $G=$ length of the  zipped file in bytes~\footnote{It may be noted that the file size produced by {\it Gzip} is rounded off to the nearest byte. Hence file sizes of 13, 15 and 16 bits will all be indicated as 2 bytes long.}
\item NSRPS measure $N =$  Number of iterations required by NSRPS algorithm to transform the given sequence into a constant sequence.
\end{itemize}

We follow a similar procedure as in~\cite{LangTree} for finding relative distance between sequences. Let the complexities of the learning sequences be $C_A$ and $C_B$. After appending the short sequence $x$ of unknown complexity, let the complexities be $C_{A+x}$ and $C_{B+x}$. If $[C_{A+x}- C_A]$ is lesser (greater) than $[C_{B+x}- C_B]$, then we classify $x$ as having the same complexity as sequence $A$ ($B$). Analysis was done with learning sequences of length 200, 100, 75, 50 and 25 and a range of test sequences varying in length from 10\% to 100\% of the original learning sequence length. We have used this procedure for all the three complexity measures mentioned above.

%
%

Tables~\ref{tab:orig_200}--\ref{tab:orig_25} summarize the results obtained for the three different measures of complexity.
Fifty trial runs were performed to come up with this result. The values in the table gives the success percentage for the measure (for every trial, if the measure is  able to identify the sequence correctly, then it is defined as a {\it success}). To ensure complete independence of all the trial runs, the initial condition for each of the fifty trials was chosen uniformly at random from the interval $[0, 1)$. \\


\begin{table}[!h]
\begin{center}
$Here \ AL \rightarrow Test ~Sequence \ Length, $
$ \ H \rightarrow \ Entropy \ measure~in~bits/symbol, $
$ \ G \rightarrow   Zipped \ length \ using \ {\it Gzip}~in~bytes,$
$ \ N \rightarrow NSRPS \ Measure.$

\caption{Results for 200 Length learning sequence. All results are given in terms of success percentage in classifying unknown sequence across 50 trials.}
\label{tab:orig_200} \vskip 5pt

\begin{tabular}{|c||c|c|c||c|c|c||c|c|c|}
  \hline
   &   \multicolumn{3}{c||}{$a=3.84$}    &  \multicolumn{3}{|c||}{$a=4.0$}  & \multicolumn{3}{|c|}{$Random$}   \\
   \hline
  AL & H & G & N & H & G & N & H & G & N \\
    & (\%) & (\%) & (\%) & (\%) & (\%) & (\%) & (\%) & (\%) & (\%) \\
     \hline
  5    & 0 & 4 & 0 & 26 & 0 & 18 & 98 & 0 & 66\\
     \hline
  25   & 0 & 55 & 10 & 28 & 0 & 92 & 82 & 2 & 94\\
     \hline
  50   & 2 & 70 & 28& 36 & 2 & 100 & 76 & 6 & 100 \\
     \hline
  80   & 16 & 92 & 44 & 42 & 2 & 100 & 62 & 6 & 100\\
     \hline
  100   & 10 & 92 & 54 & 30 & 6 & 100 & 56 & 6 & 100\\
     \hline
  135   & 10 & 92 & 70 & 42 & 38 & 100 & 62 & 10 &100 \\
     \hline
  175   & 8 & 80 & 60 & 40 & 94 & 100 & 60 & 10 & 100\\
     \hline
  200   & 8 & 66 & 52 & 42 & 100 & 100 & 50 & 10 & 100\\
     \hline

\end{tabular}
\end{center}
\end{table}

\begin{table}[!h]
\begin{center}
\caption{Results for 100 Length Learning Sequence}
\label{tab:orig_100} \vskip 5pt

\begin{tabular}{|c||c|c|c||c|c|c||c|c|c|}
  \hline
   &   \multicolumn{3}{c||}{$a=3.84$}    &  \multicolumn{3}{|c||}{$a=4.0$}  & \multicolumn{3}{|c|}{$Random$}   \\
   \hline
  AL & H & G & N & H & G & N & H & G & N \\
    & (\%) & (\%) & (\%) & (\%) & (\%) & (\%) & (\%) & (\%) & (\%) \\
     \hline
  5    & 0 & 6 & 32 & 100 & 0 & 54 & 0 & 2 & 10\\
     \hline
  15   & 2 & 22 & 36 & 90 & 6 & 78 & 8 & 6 & 60\\
     \hline
  30   & 18 & 66 &48  &66  &6  & 90 & 12 & 8 & 82\\
     \hline
  45   & 24 & 82 & 64 & 84 & 10 & 90  & 18 & 12 & 96\\
     \hline
  60   & 18 & 84 & 60 & 86 & 6 & 96 & 12 & 14 & 98\\
     \hline
  75   & 16 & 86 & 60 & 82 & 8 & 96 & 18 & 8 & 100\\
     \hline
  90   & 16 & 82 & 62 & 78 & 6 & 96 & 26 & 12 & 100\\
     \hline
  100  & 12 & 88 & 48 & 74 & 6 & 96 & 14 & 10 & 100\\
     \hline

\end{tabular}
\end{center}
\end{table}

\begin{table}[!h]
\begin{center}
\caption{Results for 50 Length Learning Sequence}
\label{tab:orig_50} \vskip 5pt

\begin{tabular}{|c||c|c|c||c|c|c||c|c|c|}
  \hline
   &   \multicolumn{3}{c||}{$a=3.84$}    &  \multicolumn{3}{|c||}{$a=4.0$}  & \multicolumn{3}{|c|}{$Random$}   \\
   \hline
  AL & H & G & N & H & G & N & H & G & N \\
    & (\%) & (\%) & (\%) & (\%) & (\%) & (\%) & (\%) & (\%) & (\%) \\
     \hline
  5    & 32 & 2 & 64 & 84 & 0 & 10 & 0 & 0 & 24\\
     \hline
  10   & 28 & 16 & 78 & 68 & 6 & 48 & 0 & 10 & 48\\
     \hline
  15   & 24 & 28 & 74 & 72 & 4 & 62 & 24 & 10 & 64\\
     \hline
  25   & 28 & 54 & 82 & 70 & 10 & 72 & 24 & 10 & 82\\
     \hline
  35   & 36 & 68 & 76 & 62 & 2 & 84 & 22 & 16 & 92\\
     \hline
  40   & 30 & 68 & 84 & 64 & 4 & 80 & 20 & 14 & 92\\
     \hline
  50   & 36 & 62 & 80 & 56 & 8 & 90 & 40 & 20 & 94\\
     \hline

\end{tabular}
\end{center}
\end{table}

\begin{table}[!h]
\begin{center}
\caption{Results for 25 Length Learning Sequence}
\label{tab:orig_25} \vskip 5pt

\begin{tabular}{|c||c|c|c||c|c|c||c|c|c|}
  \hline
   &   \multicolumn{3}{c||}{$a=3.84$}    &  \multicolumn{3}{|c||}{$a=4.0$}  & \multicolumn{3}{|c|}{$Random$}   \\
   \hline
  AL & H & G & N & H & G & N & H & G & N \\
    & (\%) & (\%) & (\%) & (\%) & (\%) & (\%) & (\%) & (\%) & (\%) \\
     \hline
  5    & 0 & 2 & 38 & 84 & 0 & 22 & 0 & 0 & 24\\
     \hline
  10   & 0 & 2 & 60 & 84 & 0 & 48 & 0 & 0 & 38\\
     \hline
  15   & 0 & 8 & 70 & 88 & 2 & 60 & 0 & 2 & 22\\
     \hline
  20   & 0 & 12 & 64 & 90 & 2 & 64 & 0 & 2 & 30\\
     \hline
  25   & 0 & 24 & 74 & 92 & 8 & 58 & 0 & 10 & 42\\
     \hline

\end{tabular}
\end{center}
\end{table}

%
%
%

%

\subsection{Analysis of results}

From the given tables, we can infer the following:
\begin {itemize}
\item For a long learning sequence (length = 200), {\it Gzip} identifies periodic sequences better while chaotic and random sequences are identified much better by NSRPS measure.
\item For intermediate length learning sequence (length = 100), {\it Gzip} identifies periodic sequences better in the case of long test sequences, while NSRPS performs better for smaller test sequences. While considering chaotic and random sequences, NSRPS outperforms the other two measures for all test sequences.
\item As we reduce the learning sequence lengths to 50 and 25, it is seen that NSPRS clearly identifies periodic, chaotic and random sequences while the other two fail to do so. The only exception in this case is when entropy measure identifies chaotic sequence better than NSRPS measure for a learning sequence of length 25.
\end {itemize}

\section{Conclusions and Future Work}

There are several measures of complexity among which Shannon's entropy is one good measure. This measure is used for automatic classification of sequences of unknown complexity given longer sequences of known complexities. However, estimating entropy measure either by use of equation (1) or by means of lossless data compression algorithms such as LZ-77 has drawbacks. One of the main drawbacks is the poor classification performance for short sequences. 

It is observed that {\it Gzip} measure doesn't achieve high accuracy for identifying chaotic and random sequences, while it works with medium efficiency for periodic sequences with a learning sequence of lengths greater than 50. This is due to the fact that {\it Gzip} uses a learning based algorithm and due to the repetitive nature of periodic sequences, learning can be done much sooner. Entropy measure (using equation (1) directly for computation) on the other hand has good efficiency while identifying chaotic sequences but performs poorly for periodic and random sequences. NSRPS is found to be very efficient in dealing with long, intermediate and short sequences.\\

Future work can be along the following lines:

\begin {itemize}
\item Classify sequences from several chaotic dynamical systems such as Tent map, Skew-tent map, binary map, Lozi map, Henon map etc. Extend this method for sequences from flows (continuous dynamical systems such as Lorenz, Rossler etc.).
\item Study the effect of noise added to learning and test sequences and analyze the performance of the complexity measures for classifying such noisy sequences.
\item Investigate the use of NSRPS measure for language classification and authorship identification applications.
\item Analyze the use of NSRPS as a lossless compression algorithm for short sequences and compare with standard compression algorithms like Huffman, Shanon-Fano, Arithmetic coding and LZW.
\end {itemize}

\section*{Acknowledgments}
\noindent {\bf{Acknowledgments:}} The authors express their
heart-felt gratitude to Mata Amritanandamayi Devi (affectionately
known as `Amma' which means `Mother') for her constant support in
material and spiritual matters. Nithin Nagaraj thanks Sutirth Dey (IISER, Pune) for useful discussions. The authors thank Dept. of Biotechnology, Govt. of India for funding through the RGYI scheme.

\end{document}